\documentclass[11pt,letterpaper,authoryear]{article}
\usepackage[english]{babel}
\usepackage{amsmath,amsthm}
\usepackage{amsfonts}
\usepackage{amssymb}
\usepackage{hyperref,hypernat}
\def\1{\mbox{1\hspace{-.35em}1}} 
\def\R{\mathbb{R}}
\def\N{\mathbb{N}}
\def\P{\mathbb{P}}
\def\E{\mathbb{E}}
\def\L{\mathbb{L}}

\def\Z{\mathbb{Z}}
\def\var{\mbox{Var\,}}
\def\lip{\mbox{Lip\,}}
\def\Lip{\mbox{Lip\,}}
\def\cov{\mbox{Cov}}
\def\Cov{\mbox{Cov}}

\def\eqsp{\;}
\def\rme{\mathrm{e}}
\def\rmi{\mathrm{i}}

\def\limiteloi{\renewcommand{\arraystretch}{0.5}
\begin{array}[t]{c}\stackrel{{\cal D}}{\longrightarrow} \\
{\scriptstyle N\rightarrow \infty} \end{array}
\renewcommand{\arraystretch}{1}}

\def\limiteas{\renewcommand{\arraystretch}{0.5}
\begin{array}[t]{c}\stackrel{a.s.}{\longrightarrow} \\
{\scriptstyle N\rightarrow \infty} \end{array}
\renewcommand{\arraystretch}{1}}

\def\at{\string@}

\newtheorem{theo}{Theorem}

\newtheorem{prop}{Proposition}
\newtheorem{Def}{Definition}
\newtheorem{cor}{Corollary}

\def\hD{{\hat d}}

\theoremstyle{remark} \newtheorem{rem}{Remark}

\begin{document}

\title{Estimation of the autocovariance function with missing observations.}
\author{Natalia Bahamonde$^{a}$, Paul Doukhan$^{b}$ and Eric Moulines$^{c}$}
\date{\today }
\maketitle

\begin{center}
{\footnotesize
$^{a}$ Instituto de Estad\'istica, Pontificia Universidad Cat\'olica de Valpara\'iso. 
\\
$^{b}$ CNRS UMR 8088 ``Analyse, G\'eom\'etrie et mod\'elisation'', Universit\'e de Cergy-Pontoise.
\\
$^{c}$ GET/T\'el\'ecom Paris, CNRS LTCI.}
\end{center}

\begin{abstract}
We propose a  novel estimator of the autocorrelation function in presence of missing observations.
We establish the consistency, the asymptotic normality, and we derive deviation bounds  for various classes of weakly dependent stationary  time series,
including causal or non causal models.  In addition, we introduce a modified version periodogram defined from these autocorrelation estimators and derive asymptotic distribution
of linear functionals of this estimator.
\end{abstract}

\section{Introduction}
The estimation of the sample autocorrelation function (hereafter ACF) from observations of $X_{1},  \dots,  X_{N}$ is important to understand the process and allows model identification.

In the classical time series analysis, the innovations  $(\epsilon_{i})_{i\in \Z}$ in the linear process $(X_{i})$ are often assumed to be independent and identically distributed
(iid), see for example \cite{BroDav91}, \cite{BoxJen}. In this case asymptotic properties of the partial sums, especially the sample ACF and the ratio of the sample covariance have been extensively studied  in the literature. A summary of results about the asymptotical theory of the sample ACF of autoregressive processes can be found for instance in \cite{BroDav91}, Chapter 7.2 and 13.3, or \cite{EmbKluMik97}, Chapter 7.3.

In practice, however, frequently the time series are not fully observed, and there may often be substantial numbers of missing values for a variety of reasons. The analysis of irregularly observed time series is one of the most important problems faced by applied researchers whose data arise in the form of time series. The study of the asymptotic properties of the ACF function of a time series model in presence of missing observations is more difficult than in the complete case.

Most of the literature above asymptotic properties of time series with missing observations is concerned with linear processes with normal innovations. In addition, these perturbations are usually regarded as strict white noise. This assumption is very restrictive; this characteristic implies only linear models with homoskedastic conditional variances. As far as we know, the first study that extended the sample ACF to the case of missing observations is \cite{Par63}. Their study formulated that the values of the observed series at unequally spaced times can be represented as an amplitude modulated time series $Y_{i}=C_{i} X_{i}$ where $(C_{i})_{i\in \Z}$ represents the censoring process. The asymptotic properties of this modified ACF were investigated in \cite{DunRob81b} under various assumptions on the noise of the linear representation $(\epsilon_{i})_{i\in \Z}$. More recently, \cite{YajNis99} compare three estimators of the autocorrelation function for a stationary process with missing
observations. The first estimator is the sample ACF extended to the case with censored data proposed originally by \cite{Par63}. The others estimators are extensions of this first estimator. The authors derive asymptotic distribution for both short memory and long memory models for the three estimators of the ACF with missing observations. They impose the same assumptions on the innovations  $(\epsilon_{i})_{i\in \Z}$ as those in \cite{DunRob81b}.

The results obtained for the weak convergence studies for sample ACF in presence of missing observation assume asymptotic stationary to fourth order for the $(C_{i})_{i\in \Z}$, then the central limit theorem is given by \cite{DunRob81b} for $\sqrt{N}(\hat{\gamma}_{Y, N}(\ell)- \hat{\gamma}_{C, N}(\ell) \gamma_{X}(\ell))$. From this, the central limit theorem can be deduced for the $\sqrt{N}(\tilde{\gamma}_{X, N}(\ell)-\gamma_{X}(\ell))$ or
$\sqrt{N}(\tilde{\rho}_{X, N}(\ell) - \rho_{X}(\ell))$, where
$\tilde{\rho}_{X, N}(\ell)=\tilde{\gamma}_{X, N}(\ell)/\tilde{\gamma}_{X, N}(0)$ is the lag-$\ell$ serial correlation and
$\rho_{X}(\ell)=\gamma_{X}(\ell)/\gamma_{X}(0)$. In particular, if the $(X_{i})_{i\in \Z}$ are iid with finite fourth moment then
$\tilde{\rho}_{X, N}(\ell)$ are asymptotically independent normal.

The asymptotic problem of the sample ACF becomes more difficult if dependence among  $(\epsilon_{i})_{i\in \Z}$ is allowed. Financial
time series often exhibit that the conditional variance can change over time, namely heteroskedasticity. Thus, the classical
limit theorems  cannot be directly applied to process with the above condition.

Theorem 6.7 in \cite{HalHey80} (p. 188) asserts asymptotic normality of sample correlations for martingale differences $(\epsilon_{i})_{i\in \Z}$ for which $\E(\epsilon_{i}^{2}|\mathcal{F}_{i-1})$ = a positive constant. In the literature the above condition is widely used. However, this condition appears too restrictive and it excludes many important models. Among them the most interesting case is the ARCH model. Thus, limit theorems by \cite{HalHey80} or \cite{DunRob81b} cannot be directly applied to linear processes with ARCH innovations. Our results avoid this limitation.

On the other hand, various generalizations of independence have been introduced in order to extend the
theory that exists to the independence framework to the more general models. The more recent is the notion of weak dependence
introduced and developed by \cite{DL}. Our choice is explained by numerous reasons; the frame of weak dependence
includes large classes of models and can be easily used in a very large statistic problems.

We shall consider the estimator of the ACF in presence of missing observations. The asymptotic behavior of the sample ACF is examined for a very general process included for the first time process whose innovations are dependent. Central limit theorems are established under fairly mild conditions.

Two frame of weak dependence are considered in this study. The first one exploits a causal property of dependence, the $\theta$-weak
dependence property (see \cite{DedDou03}). Under some conditions, the asymptotic normality of the covariance function
with missing or censored observations is found. The second frame of weak dependence, the $\lambda$-weak dependence property (see
\cite{dw}), which includes $\eta$ and $\kappa$-weak dependences. This notion is convenient for Bernoulli shifts with
associated inputs.

The paper is organized as follows. In Section 2 we introduce the notation and various weak dependent coefficients. Section 3
is devoted to limit theorems for causal and non causal weakly dependent time series. Proofs and technical results are given in the last section.

\section{Notations and Main assumptions}
\label{sec:prel}

Let $(X_{i})_{i\in \Z}$ be a discrete-time second-order stationary time series with (zero-mean). Following \cite{Par63}, we assume that the observations are given by
\begin{equation}\label{eq:replace}
Y_i=C_i X_i,
\end{equation}
where $(C_i)_{i \in \Z}$ is a non-negative modulating process taking values in $[0,1]$. When $C_i$ takes values in $\{0,1\}$, the observations are censured, but more general modulations can be considered
as well. Throughout the paper, this process is assumed to be independent from   $(X_{i})_{i\in \Z}$. This property is essential in order to allow recovery of the covariance structure of $(X_i)_{i \in \Z}$.

We denote by $\overline{X}_{N}, \overline{Y}_{N}$ the sample means of $(X_i)_{i=1}^n$ and $(Y_i)_{i=1}^n$ and by $\hat{\gamma}_{X, N}(\ell)$ and $\hat{\gamma}_{Y, N}(\ell)$
the usual estimates of the covariances $\gamma_{X}(\ell)=\mathrm{Cov}(X_{0},X_{\ell})$ and $\gamma_{Y}(\ell)=\mathrm{Cov}(Y_{0},Y_{\ell})$.

  The so-called Parzen estimator of the autocovariance coefficient $\gamma_X(\ell)$, is given by
  \begin{equation}
  \label{eq:parzenEstimator}
  \tilde{\gamma}_{X, N}(\ell)= \frac{\sum_{i=1}^{N-\ell} (Y_i - \bar{Y}_N)(Y_{i+\ell} - \bar{Y}_N)}{\sum_{i=1}^{N-\ell} C_i C_{i+\ell}} = \frac{\hat{\gamma}_{Y, N}(\ell)}{\hat{\nu}_{C, N}(\ell)}  \eqsp.
  \end{equation}
 Similarly the autocorrelation function $\rho_{X}(\ell)$ is estimated by $\tilde{\rho}_{X, N}=\tilde{\gamma}_{X, N}(\ell)/\tilde{\gamma}_{X, N}(0)$.

We study both the consistency and asymptotic normality  of the autocovariance and autocorrelation functions of time series with missing observations, and also establish non-asymptotic deviation bounds. These results are obtained under general dependence structures.

\subsection{Weak-dependence measures}
\label{sec:dep}

Let $p$ be a positive integer. For $f: \R^{p}  \to  \R$ a function, define $\lip f$ the Lipschitz coefficient by:
\begin{equation*}
\lip f = \sup_{(x_{1}, \dots , x_{p}) \neq (y_{1}, \dots , y_{p})} \frac{|f(x_{1}, \dots, x_{p})-f(y_{1}, \dots,
y_{p})|}{|x_{1} - y_{1}| + \dots + |x_{p}-y_{p}|}.
\end{equation*}

\begin{Def} \cite{DL}
The vector-valued ($d \times 1$) process $(Z_{i})_{i \in \Z}$ is said to be \emph{weakly dependent} if
\begin{equation}
|\cov \left(f(Z_{s_{1}}, \dots, Z_{s_{u}}), g(Z_{t_{1}}, \dots, Z_{t_{v}}) \right)|  \leq \psi(u, v, \lip f, \lip g) \epsilon_{Z}(r),
\end{equation}
for any  real valued functions $f$ and $g$ defined respectively on $\R^{ud}$ and $\R^{vd}$, that satisfy $\|f
\|_{\infty},  \|g \|_{\infty} \leq 1$ and $\lip f, \lip g < \infty$, and for any $r \geq 0$ and any $(u + v)$-tuples such that  $s_{1} \leq \dots \leq s_{u} \leq s_{u}+r \leq t_{1} \leq \dots \leq
t_{v}$. Here, the sequence $(\epsilon_{Z}(r))_{r=0}^\infty$ is assumed to decrease to zero at infinity and  $\psi : \N^{2} \times (\R^{+})^{2}  \to  \R^{+}$ is a function.
\end{Def}

Specific functions $\psi$ yield  different notions of weak dependence which have been shown to be appropriate to cover various time-series settings \cite{bouquin}:
\begin{itemize}
\item $\psi(u, v, a, b) = vb$  corresponds to the notion of $\theta$-dependence.
\item $\psi(u, v, a, b) =uvab$, corresponds to the notion of $\kappa$-dependence.
\item $\psi(u, v, a, b) = uvab +ua + vb$, corresponds  to the notion of $\lambda$-dependence.
\end{itemize}
For simplicity the sequence $\epsilon_X(r)$ will be denoted respectively as  $\theta_X(r)$,  
$\kappa_X(r)$, and $\lambda_X(r)$.\\
 We shall also consider strong mixing coefficients $\alpha_X(r)$ related with $\psi(u,v,a,b)\equiv1$; in this case heredity is complete through measurable images, see \cite{Rio} or \cite{D}.\\
The following simple lemma, which relies on the decomposition of the covariance of two random variables conditioned to two independent $\sigma$-algebra, is intrumental in the sequel:
\begin{prop}
\label{prop:composition-processus}
Assume that $(U_i)_{i \in \Z}$ and $(V_i)_{i \in\Z}$ are two vector-valued independent processes. Assume in addition that these two processes are $(\epsilon,\psi)$-weakly dependent with the
same $\psi$-function and sequences denoted $(\epsilon_U(r))_{r \in \Z}$ and $(\epsilon_V(r))_{r \in \Z}$, respectively.
Then the vector-valued process $(W_i)_{i \in \Z}$ with $W_i= (U_i,V_i)^T$ is also $(\epsilon_W,\psi)$-weakly dependent with $\epsilon_W(r)= \epsilon_U(r) + \epsilon_V(r)$.
\end{prop}\noindent
As a consequence, provided that $(X_i)_{i \in \Z}$ and $(C_i)_{i \in \Z}$ are both $(\epsilon,\psi)$-weakly dependent processes, then the process $2\times1-$process $Z_i=(X_i,C_i)$ also:
weak dependences of its coordinates are equivalent to that of this process.
These coefficients have some hereditary properties. For example, heredity through  Lipschitz functions is clear and this may be extended to locally Lipschitz functions  \cite{bouquin}.
\begin{prop}
\label{prop:heredh}
Let $(U_n)_{n\in \Z}$ be a sequence of $\R^k$-valued random
variables. Let $m>1$. We assume that there exists some constant
$C>0$ such that $\max_{1 \leq i \leq k}\|U_i\|_m \leq C$. Let $h$ be
a function from $\R^k$ to $\R^d$ such that  $h(0)=0$ and
 for $x,y \in \R^k$, there exist $a$ in $[1,m[$ and $c>0$ such that $$|h(x)-h(y)|
\leq c  |x-y| (1+|x|^{a-1}+|y|^{a-1})\; .$$ We define the sequence
$(V_n)_{n\in \Z}$ by $V_n=h(U_n)$. Then, if $(U_n)_{n \in \Z}$ is weakly dependent then $(V_n)_{n \in Z}$ is also weakly dependent and,
\begin{itemize}
\item  $\theta_V(r) = {\cal O}\left( {\theta_U(r)}^{\frac{m-a}{m-1}}
\right) ;$

\item $\kappa_V(r)= {\cal O}\left(
{\kappa_U(r)}^{\frac{m-a}{m+a-2}} \right)  ;$

\item $\lambda_V(r)= {\cal O}\left(
{\lambda_U(r)}^{\frac{m-a}{m+a-2}} \right)  .$

\end{itemize}
\end{prop}
\begin{rem}For $U_n=(X_n,X_{n+\ell},C_n,C_{n+\ell})$
the function $h(x,x',c,c')=cc'\{xx'-\gamma_X(\ell)\}$ satisfies the previous assumptions with
$a=2$. \end{rem}

\section{Limit theorems}

In this section, we study the asymptotic properties of  the Parzen estimator of the ACF under the different dependence conditions mentioned above.
Denote by
$$\nu(\ell) = \E[C_0 C_\ell] \quad \text{and}  \quad m(\ell,k,m) = \E(C_{0}C_{\ell}C_{k}C_{m}) \eqsp. $$
As a consequence of Slutsky lemma and the results in Dedecker {\it et} alii (2007) we immediately derive

\begin{theo}
\label{theo:CLT-covariance}
Let  $(X_i)_{i\in \Z}$ be a real valued, stationary sequence time series of square integrable observed, with censored data. Let the modulating process $(C_i)_{i\in \Z}$ be
a nonnegative bounded stationary process. We assume that $(C_i)_{i\in \Z}$ be independent of the process $(X_i)_{i\in \Z}$. Assume either

\begin{itemize}
\item $(X_i)_{i\in \Z}$ and $(C_i)_{i \in \Z}$ are strong mixing stationary time series, $\E |X_{0} |^{m} < \infty $ for $m>4$ and  $\sum_{i\geq 0} i^{\frac 1{m-4}}  \alpha(i)  < \infty$.
\item $(X_i)_{i\in \Z}$ and $(C_i)_{i \in \Z}$ are $\theta$-weakly dependent stationary time series, $\E |X_{0} |^{m} < \infty $ for $m>4$ and  $\sum_{i\geq 0} i^{\frac 1{m-4}}  \theta^{\frac{m-2}{m-1}}(i)  < \infty$.
\item $(X_i)_{i\in \Z}$ and $(C_i)_{i \in \Z}$ are $\kappa$-weakly dependent stationary time series,  $\E |X_{0} |^{m} < \infty $ for $m>4$, and
$\kappa(r)= \mathcal{O}(r^{-\kappa})$ as $r \to \infty$ for $\kappa>\frac m{m-2}\Big(2 + 1/(m-2)\Big)$.
\item $(X_i)_{i\in \Z}$ and $(C_i)_{i \in \Z}$ are $\lambda$-weakly dependent stationary time series, $\E|X_{0}|^{m}< \infty$ for some $m>4$, and
$\lambda(r)= \mathcal{O}(r^{-\lambda})$ as $r \to \infty$ for $\lambda>\frac m{m-2}\Big(4 + 1/(m-2)\Big)$.
\end{itemize}
Then, under any of these assumptions,
\begin{equation*}
\sqrt{N} \nu(\ell) \big(\tilde{\gamma}_{X, N}(\ell)-\gamma_{X}(\ell) \big) \limiteloi N(0, \sigma^{2}_{\ell}) \eqsp,
\end{equation*}
where
\begin{equation}
\sigma^{2}_{\ell}= \sum_{k\in \Z} m(\ell,k,k+\ell) [\kappa_{4}(\ell, k, k+\ell) + \gamma_{X}(k+\ell)\gamma_{X}(k-\ell)-\gamma_{X}^{2}(\ell)].\label{eq:VAteo}
\end{equation}
\end{theo}

\begin{rem}
For  non-causal dependent sequences ($\lambda$ and $\kappa$--weak dependence cases), the assumptions need to be more restrictive and stronger than for the causal $\theta$--weak dependence case. For this case, indeed, the dependence condition rewrites $\theta(r)= \mathcal{O}(r^{-\theta})$ as $r \to \infty$ for $\theta>\frac {m-1}{m-2}\Big(1 + 1/(m-2)\Big)$.
\end{rem}

\begin{rem} From \cite{bouquin}, we know that conditions of Theorem \ref{theo:CLT-covariance} are sufficient to obtain the weak invariance principle in the $\theta$-weak dependence frame. Analogously, \cite{dw} show the same principle in the $\lambda$ and $\kappa$--weak dependence cases.
\end{rem}
By polarization, we simply derive the following extension of this theorem which will be used hereafter (the proof is left to the reader).
\begin{cor}
\label{cor:CLT-covariance}
Under the assumptions of Theorem \ref{theo:CLT-covariance}, for all $k \in \N,  \ell_{1}< \dots < \ell_{k} \in \N^{k}$,
\begin{equation*}
\sqrt{N} \biggl(\nu(\ell_{i}) \big(\tilde{\gamma}_{X, N}(\ell_{i})-\gamma_{X}(\ell_{i}) \big)\biggl)_{1 \leq i \leq k} \limiteloi N_{k}(0, \Sigma(\ell_{1}, \dots, \ell_{\ell}))
\end{equation*}
where $\Sigma(\ell_{1}, \dots, \ell_{k})=\big( \sigma_{\ell_{i}\ell_{j}}\big)_{1\leq i,j\leq \ell_{k}}$ is defined by
\begin{align}
&\sigma^{2}_{\ell_{i}, \ell_{j}} \label{eq:VAjoint} \\ \nonumber
&\; = \sum_{k\in \Z} m(\ell_i,k,k+\ell_j) [\kappa_{4}(\ell_{i}, k, k+\ell_{j}) + \gamma_{X}(k+\ell_{j})\gamma_{X}(k-\ell_{i})-\gamma_{X}(k)\gamma_{X}(k+\ell_{j}-\ell_{i})] \eqsp.
\end{align}
\end{cor}
The following results provide {\it a.s.} asymptotic behaviour of the Parzen autocovariance.
\begin{theo}
\label{theo:SLLN:covariance}
\begin{equation*}
\tilde{\gamma}_{X, N}(\ell)-\gamma_{X}(\ell)  \limiteas 0 \eqsp,
\end{equation*}
if $\E|X_{0}|^{m} < \infty$ for some $m=2+\delta$  and, either the processes are $\theta$-dependent with  $\sum_{i\geq 0} i^{\frac{r(\delta-1)+1}{r-(1+\delta)}}  \theta^{\frac\delta{\delta+1}}(i)  < \infty \mbox{ for some }\delta>0$, or they are $\kappa,$ or $\lambda$-dependent and satisfy assumptions from theorem \ref{theo:CLT-covariance}.
\end{theo}


\section{Division}
\label{sec2}
Let $(U_i,V_i)_{i\in\Z}$ be a stationary
sequence and set $\displaystyle\widehat D_n= 1/n\sum_{i=1}^nU_i$,
$\displaystyle\widehat N_n= 1/n\sum_{i=1}^nU_iV_i$ then $N_n=N=\E
U_1V_1$, $D_n=D=\E U_1$ and $\displaystyle \widehat R_n=\widehat
N_n/\widehat D_n,$ $R_n=R={N}/{D}.$

\begin{theo}\label{wsums}
Let $(U_i,V_i)_{i\in\Z}$ be a stationary sequence with $U_i\ge 0$
{\it (as.)}. Let $0<p<q$  and assume that for $\displaystyle
r=\frac{pq}{q-p}$ and $\displaystyle
s=\frac{p(q+2)}{q-p}$:$$\|U_iV_i\|_r\le c ,\qquad \|V_i\|_{s}\le
c.$$Assume that the dependence structure of the sequence
$(U_i,V_i)_{i\in\Z}$ is such that
\begin{equation}\label{NDracn}
\big\|\widehat D_n -D\big\|_q \leq \frac C{\sqrt n}, \quad
\big\|\widehat N_n -N\big\|_p \leq \frac C{\sqrt n}
\end{equation} then
$\displaystyle \|\widehat R_n- R\|_p={\cal O}\left( 1/{\sqrt
n}\right).$
\end{theo}
In the following cases, we assume that $\|V_i\|_{s}\le c$ and $
\|U_iV_i\|_r\le c$  and prove that (\ref{NDracn}) holds. Denote
$Z_i=U_iV_i-\E U_iV_i $. For simplicity we will often assume
$\|U_i\|_\infty<\infty$, $\|V_i\|_r<\infty$.


\subsection{Independent case}
Assume that $(U_i,V_i)$ is i.i.d.  Assume that $ \|U_0\|_q\le c$
and $\|U_0V_0\|_p\le c$. From the Marcinkiewikz-Zygmund inequality
for independent variables, for $2\leq q\leq r$, we get
\begin{eqnarray*}
  \E\left|\widehat
D_n -D\right|^q &\leq& C_q  \E |U_1|^q n^{-\frac q2}
  \, \leq\, C  n^{-\frac q2}, \\
    \E\left|\widehat N_n -N\right|^p& \leq &C_p\E
|U_1V_1|^p n^{-\frac p2}
  \, \leq \,C  n^{-\frac p2},
\end{eqnarray*}
and  (\ref{NDracn}) holds. Now H\"older inequality implies those
relations if $\|U_0\|_q,$ and $\|V_0\|_{\frac{qp}{q-p}}<\infty.$


\subsection{Strong mixing case}
 Denote $(\alpha_i)_{i\in \N}$ the strong mixing
coefficient sequence of the stationary sequence $(U_i,V_i)_{i\in
\N}$.
\begin{prop}\label{wsummix}Assume that for $r'>q$, $\|U_0\|_{r'}\le c$.
Relation (\ref{NDracn}) holds if $\alpha_i={\cal O}(i^{-\alpha})$
with
   $$\alpha >
   \left(\frac{p}2\,\cdot\frac{r}{r-p}\right)
   \vee\left(\frac{q}2\,\cdot\frac{r'}{r'-p}\right),
   $$
\end{prop}


\subsection{Causal weak dependence}
 Define the
$\gamma$ coefficient of dependence of a centered sequence
$(W_i)_{i\in \N}$ with values in $\R^d$ by
$$
\gamma_i=\sup_{k\geq 0}\left\|\E(W_{i+k}|\,{\cal M}_k)\right\|_1
$$
\begin{prop}\label{wsumwd}Assume that $\|U_0\|_\infty\le c$. Relation (\ref{NDracn}) holds if the sequence of coefficients $\gamma$
 associated to the stationary sequence
$(W_i)_{i\in \N}=(U_i,V_i)_{i\in \N}$ is such
   that $\gamma_i={\cal
O}(i^{-\gamma})$ with
   $$\gamma >
   \left(\frac{p}2\,\cdot\frac{r-1}{r-p}\right)
   \vee\frac{q}2
   .$$
\end{prop}


\subsection{Non causal weak dependence}
Here we consider  non causal weakly dependent stationary sequences
of bounded variables and assume that $q$ and $p$ are integers. A
sequence $(W_i)_{i\in \N}$ is said to be $\lambda$-weakly
dependent if there exists a sequence $(\lambda(i))_{i\in\N}$
decreasing to zero at infinity such that:
\begin{equation*}
\Big|\Cov \Big(g_1(W_{i_1},\ldots,W_{i_u}), g_2(W_{j_1},\ldots,W_{j_v})\Big) \Big|
 \leq \left(u\Lip g_1+v  \Lip g_2+uv\Lip g_1 \Lip g_2  \right)  \lambda(k),
\end{equation*}
for any $u$-tuple $(i_1,\ldots,i_u)$ and any $v$-tuple
$(j_1,\ldots,j_v)$ with $i_1\leq \cdots \leq i_u < i_u+k\leq
j_1\leq\cdots\leq j_v$ where $g_1,g_2$ are two real functions of
$\Lambda^{(1)}=\{g_1\in \Lambda|\,\|g_1\|_\infty\le1\}$
respectively defined on $\R^{Du}$ and $\R^{Dv}$ ($u,v\in \N^*)$.
Recall here that $\Lambda$ is the set of functions with $\Lip
g_1<\infty$ for some $u\ge1$, with
$$
\Lip
g_1=\sup_{(x_1,\ldots,x_u)\ne(y_1,\ldots,y_u)}\frac{\left|g_1(y_1,\ldots,y_u)-g_1(x_1,\ldots,x_u)\right|}{|y_1-x_1|+\cdots+|y_u-x_u|}.
$$
 The monograph Dedecker {\it et al.} (2007) \cite{bouquin} details weak dependence concepts, as well as extensive models and  results.
\begin{prop}\label{wsumwdnc}Assume that $p$ and $q\ge2$ are even
integers. Assume that the stationary sequence $(U_i,V_i)_{i\in
\N}$ is $\lambda$-weakly dependent. Assume that $Z_0=U_0V_0-\E
U_0V_0$ is bounded by $M$. Relation (\ref{NDracn}) holds if
$\lambda(i)={\cal O}(i^{-\lambda})$ with $\lambda>\frac{q}{2}$.
\end{prop}
\paragraph{Remarks}\begin{itemize}
    \item Unbounded random variables may also be considered under an additional concentration
    inequality ($\P(Z_i\in(x,x+y))\le Cy^a$ for some $a>0$) and  Theorem 3 and Lemma 1
    from Doukhan and Louhichi (1999) \cite{DL}, imply that the same relation holds if
$\E|Z_i|^{q+\delta}<\infty,$
 $\sup_{x,i}\P(Z_i\in(x,x+y))\le Cy^a,$  ($\forall y>0$), and
  $\displaystyle\sum_{n=1}^\infty
 n^{q\frac{q+\delta}{2\delta}-1}\lambda^{\frac
 a{2+a}}(n)<\infty$.
    \item Non-integer moments $q\in(2,3)$ are considered in Doukhan and Wintenberger (2007)
(see \cite{dw}, Lemma 4), and the same inequality holds if
$\E|Z_i|^{q'}<\infty$ with $q'=q+\delta$,
 and $\lambda(i)={\cal O}(i^{-\lambda})$ with  $\lambda>4+2/q'$ for $q$ small enough:
$$
q\le 2+\frac12\left(
\sqrt{(q'+4-2\lambda)^2+4(\lambda-4)(q'-2)-2}+q'+4-2\lambda\right)\;\big(\le
q'\big).
$$
\end{itemize}

\section{Spectral estimation}
\label{sec:spectral}

In this section we study estimation of functionals of the  spectral density function from the censored time
series. Using the Parzen estimator of the covariance  $\tilde{\gamma}_{X, N}(\ell)$, 
estimates of the spectral density of $(X_{i})_{i\in \Z}$ can be constructed.
More precisely, we introduce a modified periodogram defined with the empirical covariance $\tilde{\gamma}_{X, N}(\ell)$ of the
censored process
\begin{equation*}
\tilde{I}_{N}(\lambda) = \sum_{\ell \in \Z} \tilde{\gamma}_{X, N}(\ell) \rme^{-\rmi\ell\lambda} \eqsp.
\end{equation*}
Assume now that we wish to estimate linear functionals of the Parzen peridogram,
\begin{equation*}
J_{X}(g)=\int_{-\pi}^{+\pi}g(\lambda)f_{X}(\lambda),
\end{equation*}
where $f_{X}(\lambda)$ is the spectral density of $(X_{i})_{i \in \Z}$. Using the estimates of the covariances of $(X_{i})_{i \in \Z}$ given by $\tilde{\gamma}_{X, N}(\ell)$, estiamtes of the integrated periodogram of $(X_{i})_{i \in \Z}$ can be constructed by
\begin{equation*}
\tilde{J}_{X, N}(g) = \int_{-\pi}^{+\pi}g(\lambda) \tilde{I}_{N}(\lambda) d\lambda.
\end{equation*}
Under general conditions the integrated periodogram is a consistent estimator of the $J_{X}(g)$ provided
the spectral density $f_{X}(\lambda)$ is well defined.
\\
Our aim is to study the asymptotic behavior of $\E|\tilde{J}_{X, N}(g) - J_{X}(g)|^{q}$. We consider the Sobolev space $\mathcal{H}_{s}$
for $s>1$:
\begin{equation*}
\mathcal{H}_{s}=\{ g \in \L^{2}[ -\pi, \pi]; g(-x)= g(x), \|g\|^{2}_{\mathcal{H}_{s}} < \infty\}, \quad \mbox{with} \quad
\|g\|^{2}_{\mathcal{H}_{s}}=\sum_{\ell \in \Z} (1+|\ell|)^{S} |g_{\ell}|^{2}.
\end{equation*}
for $g$ a $2\pi$--periodic function such that $g \in \L^{2}([-2\pi, 2\pi[)$ and $g(\lambda)=\sum_{\ell \in \Z} g_{\ell}e^{i\ell \lambda}$.
\\
The norm of the dual space $\mathcal{H}'_{s}$ of $\mathcal{H}_{s}$ writes
\begin{equation*}
\|T\|^{2}_{\mathcal{H}'_{s}}=\sup_{\|g\|_{\mathcal{H}_{s}}\leq 1} |T(g)|^{2}=\sum_{\ell \in \Z} (1+|\ell|)^{-s} |T(e^{i\ell \lambda})|^{2}.
\end{equation*}

Note that $\tilde{J}_{X, N}(g)$ and $J_{X}(g) \in \mathcal{H}'_{s}$. We have

\begin{theo}\label{theo:LGNuniforme}
Let  $(X_i)_{i\in \Z}$ is a real valued time series observed with censored data such that $\sum_{\ell \in \Z} \gamma_{X}(\ell)^{2}<\infty$ and $(\kappa_{4}(i,j,k))_{i,j,k}$ the fourth cumulants of $X$ exist. If assumptions of Lemma 2 are satisfied, them $\forall g \in \mathcal{H}_{s}$
\begin{equation*}
\lim_{N\to \infty} \E \|\tilde{J}_{X, N}(g) - J_{X}(g)\|_{\mathcal{H}'_{s}}^{2} \longrightarrow  0.
\end{equation*}
\end{theo}

\begin{theo}
Under assumptions of Theorem \ref{th:espe2}, the central limit theorem is satisfied if we ssume that $\|\hat{\gamma}_{C, N}(\ell) - \gamma_{C}(\ell) \|_{q} \leq v_{n} $  for some $q>2$, and  $\|\hat{\gamma}_{Y, N}(\ell) - \gamma_{Y}(\ell) \|_{2} \leq v_{n} $.  Moreover if $\| X_{i}X_{i+\ell}C_{i}C_{i+\ell} \|_{r}\leq K$, and  $\| X_{i}X_{i+\ell} \|_{s}\leq k$. Then

\begin{equation*}
\sqrt{N}[\tilde{J}_{X, N}(g)-J_{X}(g)] \limiteloi N(0, \sigma^{2}(g)) \qquadå \mbox{ in the space } \mathcal{H}'_{s}.
\end{equation*}
\end{theo}


\section{Technical results and proofs}
\noindent
Our proof for central limit theorems is based on a weak invariance principle under weak dependence conditions.

\subsection{Proof of Theorem \ref{theo:CLT-covariance}}

To study the asymptotic behavior of $\tilde{\gamma}_{X, N}(\ell)= {\hat{\gamma}_{Y, N}(\ell)}/{\hat{\nu}_{N}(\ell)}$ we
decompose the quantity of interest:

\begin{eqnarray}
\sqrt{N} \left\{ \tilde{\gamma}_{X, N}(\ell)-\gamma_{X}(\ell) \right\} &=& \frac{\hat{\gamma}_{Y, N}(\ell)}{\hat{\nu}_{N}(\ell)} - \gamma_{X}(\ell) \nonumber\\
&=& {\hat{\nu}^{-1}_{N}(\ell)} \cdot \sqrt{N} \left\{ \hat{\gamma}_{Y, N}(\ell) - \gamma_{X}(\ell) \hat{\nu}_{N}(\ell) \right\}\label{1y2}.
\end{eqnarray}
The second factor in the RHS of the previous expression is handled by using a Central Limit Theorem for weakly dependent sequences.
\begin{equation}
\sqrt{N} \left\{ \hat{\gamma}_{Y, N}(\ell) - \gamma_{X}(\ell) \hat{\nu}_{N}(\ell) \right\}
= N^{-1/2}  \sum_{k=1}^{N-\ell} Z_k\label{zeta},
\end{equation}
with $Z_{k} = C_{k} C_{k+\ell} \left( X_{k} X_{k+\ell}-\gamma_{X}(\ell) \right) $. According to Propositions \ref{prop:composition-processus} and \ref{prop:heredh}, $(Z_k)$ is
weakly dependent. For the $\kappa$ and $\lambda$-dependence.
\cite[Theorem 7.1, Theorem 7.2, p. 154]{bouquin} prove a C.L.T.  for $\kappa$ and $\lambda$--weak dependent processes, and the proof is immediate.
The situation is a bit more intricate for $\theta$-dependence.
We shall show that $S_{N}=\sqrt{N}(\tilde{\gamma}_{X, N}(\ell)-\gamma_{X}(\ell))$ is asymptotically normally distributed by showing that $N^{-1}\sum Z_{k}$ satisfies a central limit theorem for $\theta$--weak dependent variables. Theorem 2 of \cite{DedDou03} states that if $S_{N}=N^{-1}\sum Z_{k}$ be a strictly stationary sequence of square integrable an centered random variables, then if the condition
\begin{equation}
 Z_{0}\E(S_{N} \mid \mathcal{M}_{0})\mbox{ converges in } \L_{1}; \label{eq:DR}
\end{equation}
holds,
where $\mathcal{M}_{i} = \sigma (Z_{j}; j<i)$, then $N^{-1/2}S_{N}$ will be asymptotically normally distributed.
\\
Alternatively,  by Corollary 1 of \cite{DedDou03}, in the $\theta$-weak dependence frame, $D(2; X_{0})$ implies (\ref{eq:DR}) where
\begin{equation}
D(p, X): \int_0^{\|X\|_1} (\theta^{-1} (2u))^{p-1} Q_X^{p-1} \circ G_X(u)du  < \infty.\label{eq:D}
\end{equation}
and for  $X$  a real valued random variable,  $Q_X$  denotes the generalized inverse of the tail function $x \mapsto
\P(|X|\leq x)$, and $G_X$ the inverse of $x \mapsto \int^x_0 Q_X (u) du$.
\cite[Corollary 7.6]{bouquin} gives sufficient conditions to satisfy the condition $D(p;  X )$. In particular, if $p=2$ and $\|X_{0} \|_{4+\delta}\leq \infty$ then $\|Z_{0}
\|_{2+\delta}\leq \infty$ and $\sum_{i\geq 0} i^{1/\delta}  (\theta_X(i))  < \infty$ for some $\delta>0$, \eqref{eq:DR} condition is
satisfied and therefore the asymptotic normality in \eqref{zeta} follows from \cite[Corollary 7.5]{bouquin}.
\medskip

\noindent
It remains to show that the limiting covariances are given by $\sigma^{2}_{\ell}=\sum_{k\in \Z}\E Z_{0}Z_{k}$.
\begin{equation}
\sigma^{2}_{\ell} = \sum_{k\in \Z} \E(C_{0}C_{\ell}C_{k}C_{k+\ell})[ \E(X_{0}X_{\ell}X_{k}X_{k+\ell})-3\gamma_{X}^{2}(\ell) ].\label{eq:EAV1}
\end{equation}
We use the following identity for $ (i, j, k) \in \Z^{3}$ for simplify the expression (\ref{eq:EAV1}):
\begin{equation*}
\kappa_{4}(i, j, k)=\E X_{0}X_{i}X_{j}X_{k}-\E X_{0}X_{i}\E X_{j}X_{k}-\E X_{0}X_{j}\E X_{i}X_{k}-\E X_{0}X_{k}\E X_{i}X_{j} 
\end{equation*}
Then the following expression, which exists for all finite $\ell$,  is equivalent to the asymptotic variance in equation (\ref{eq:EAV1})
\begin{equation*}
\sigma^{2}_{\ell} =  \sum_{k\in \Z} \E(C_{0}C_{\ell}C_{k}C_{k+\ell})[\kappa_{4}(\ell, k, k+\ell) + \gamma_{X}(k+\ell)\gamma_{X}(k-\ell)-\gamma_{X}^{2}(\ell)].
 \end{equation*}
We note that when all the $(X_{i})_{i \in \Z}$ are observed $\E(C_{0}C_{\ell}C_{k}C_{k+\ell})=1$ so that $\sigma^{2}_{\ell}$ agrees with \cite{Ros85}, Theorem 3, p. 58.\\
\medskip

\noindent
Finally the first factor in \eqref{1y2} converges in probability to $\nu(\ell)$ because the stated assumptions ensuring a central limit theorem also imply the convergence of the empirical variances.
Slutsky's Theorem allows to complete the proof. This completes the proof of Theorem \ref{theo:CLT-covariance}.~$\qquad\square$\\
Corollary \ref{cor:CLT-covariance} is standard from the previous result. 

\subsection{Proof of Theorem \ref{theo:SLLN:covariance}}
To obtain strong laws for the sample autocorrelation function of $(X_{i})_{i\in \Z}$ estimated like a function of $(Y_{i})_{i\in \Z}$
and $(C_{i})_{i\in \Z}$ in the $\theta$-weak dependence frame, by theorem 3 of \cite{DedDou03} if $D(p, X)$ holds for some
$p \in [1, 2[$, then $n^{-1/p}\sum_{i=1}^{n}(X_{i}-\E(X_{i}))$ converges almost surely to $0$ as $n$ goes to infinity. We use again Lemma 2 of \cite{DedDou03}, we need considered a $p >1$ and therefore, we need weaker moments conditions.
 If  $\|X_{0} \|_{r} < \infty$ for some $r>2(1+\delta)$  and  $\sum_{i\geq 0} i^{(r(\delta-1)+1)/(r-(1+\delta))}  \theta_i  < \infty \mbox{ for some }\delta>0$, then is a sufficient condition for $\tilde{\gamma}_{X, N}(\ell)-\gamma_{X}(\ell)  \stackrel{a.s.}{\longrightarrow } 0$.\\
For the strong laws for $\lambda$ or $\kappa$--weak dependence cases, \cite{dw} proved a bound
of the $(2+\delta)$-moment of the sum of a process  $\lambda$ or $\kappa$--weak dependence. This bound it directly yields the strong law of
large numbers using the Borel-Cantelli lemma.

\subsection{Proof of Theorem \ref{theo:LGNuniforme}}
In order to prove that $\|\tilde{J}_{X, N}(g) - J(g)\|_{\mathcal{H}'_{s}}^{2}$ converges, we use a bound for $\E\|\tilde{J}_{X, N}(g) - J(g)\|_{\mathcal{H}'_{s}}^{2}$ under further conditions. Then
\begin{eqnarray*}
\|\tilde{J}_{X, N}(g) - J(g)\|_{\mathcal{H}'_{s}}^{2}&=& \|\sum_{\ell\in \Z} (\tilde{\gamma}_{X, N}(\ell)-\gamma_{X}(\ell)) g_{\ell}\|^{2}_{\mathcal{H}'_{s}}\\
&=& \|\sum_{\ell\in \Z}(1+|\ell|)^{-s} (\tilde{\gamma}_{X, N}(\ell)-\gamma_{X}(\ell))  (1+|\ell|)^{s}g_{\ell}\|^{2}_{\mathcal{H}'_{s}}\\
&\leq & \biggl( \sum_{\ell\in \Z}(1+|\ell|)^{-2s} (\tilde{\gamma}_{X, N}(\ell)-\gamma_{X}(\ell))^{2}\biggl) \biggl(\sum_{\ell\in \Z} (1+|\ell|)^{2s}g_{\ell}^{2}\biggl)\\
&\leq & \|g\|_{\mathcal{H}_{2s}} \biggl( \sum_{\ell\in \Z}(1+|\ell|)^{-2s} (\tilde{\gamma}_{X, N}(\ell)-\gamma_{X}(\ell))^{2}\biggl)
\end{eqnarray*}
Therefore, to prove a bound for $\E\|\tilde{J}_{X, N}(g) - J(g)\|_{\mathcal{H}'_{s}}^{2}$ it sufficient to show that $\E(\tilde{\gamma}_{X, N}(\ell)-\gamma_{X}(\ell))^{2}$ is bounded.
\begin{equation*}
\E\|\tilde{J}_{X, N}(g) - J(g)\|_{\mathcal{H}'_{s}}^{2}\leq \|g\|_{\mathcal{H}_{2s}} \biggl( \sum_{\ell\in \Z}(1+|\ell|)^{-2s} \E(\tilde{\gamma}_{X, N}(\ell)-\gamma_{X}(\ell))^{2}\biggl)
\end{equation*}
And,
\begin{equation*}
\E(\tilde{\gamma}_{X, N}(\ell)-\gamma_{X}(\ell))^{2} \leq v_{n}\biggl( \frac{1}{\gamma_{C}(\ell)}+
\frac{K}{\gamma_{C}(\ell)^{2}} + \frac{v_{N}^{\alpha}(Nk)^{\frac{1}{s}}}{\gamma_{C}(\ell)^{\alpha+1}}\biggl),
\end{equation*}


\begin{eqnarray*}
\Big\|\max_{1\le i\le n}|X_{i}X_{i+\ell}|\cdot\frac{|\hD-d|^{1+\alpha}}{|d|^\alpha}\Big\|_2
&\le&\frac1{|d|^\alpha} \|\max_{1\le i\le n}|X_{i}X_{i+\ell}|\|_{2a}\|\hD-d|^{1+\alpha}\|_{2b}\\
&\le&\frac1{|d|^\alpha} \left(\E\max_{1\le i\le n}|X_{i}X_{i+\ell}|^{2a}\right)^{\frac1{2a}}v_n^{1+\alpha} \label{terme}
\end{eqnarray*}
if $q\ge 2b(1+\alpha)$ or equivalently $q-2(1+\alpha)\ge q/a$. We need an argument of \cite{pisier} written as
follows: assume that $\varphi:\R^+\to\R^+$ is convex and non decreasing then
\begin{equation}
\label{eq:pisier}
\varphi\left( \E \max_i |X_{i}X_{i+\ell}|^{2a}\right) \leq \E \varphi\left( \max_i |X_{i}X_{i+\ell}|^{2a}\right)\leq \E
\sum_i\varphi\left(  |X_{i}X_{i+\ell}|^{2a}\right)\le \sum_i\E \varphi\left(  |X_{i}X_{i+\ell}|^{2a}\right)
\end{equation}
Hence $\E \max_i |X_{i}X_{i+\ell}|^{2a}\le (nc)^{2a/s}$ with $\varphi(x)=x^{s/2a}$. Now the bound in the right hand side of
(\ref{terme}) can be specified as $v_n^{1+\alpha}(nk)^{\frac1s}/{|d|^\alpha}$ if $s\ge pa$ this holds because $\displaystyle
1-\frac 2q(1+\alpha)\ge \frac1a\ge \frac 2s$, if $\alpha>0$ is small enough with $\frac12\ge\frac {1+\alpha}q+\frac1s$.
We thus obtain $\gamma_C(\ell)\Big\|\frac{\hat\gamma_{Y, N}(\ell)}{\hat\gamma_{C, N}(\ell)}-\gamma_X(\ell)\Big\|_2 \leq v_{n}+ \frac K
d v_{n}+v_n^{1+\alpha}(nk)^{\frac1s}/{|d|^\alpha}$, that implies the result of the Theorem.$\qquad\square$

\subsection{Proof of Theorem \ref{theo:LGNuniforme}}

\begin{enumerate}
\item
\begin{eqnarray*}
J_{N}(g)-J(g)&=& \sum_{|\ell|<N}\tilde{\gamma}_{X, N}(\ell)g_{\ell}- \sum_{\ell \in N} \gamma_{X}(\ell)g_{\ell}\\
&=& \sum_{|\ell|<N}\tilde{\gamma}_{X, N}(\ell)g_{\ell}-\sum_{|\ell| \geq N} \gamma_{X,}(\ell)g_{\ell}-\sum_{|\ell| < N} \gamma_{X}(\ell)g_{\ell}\\
&=& -\sum_{|\ell| \geq N} \gamma_{X}(\ell)g_{\ell}-\sum_{|\ell| < N}(\tilde{\gamma}_{X}(\ell)-\gamma_{X}(\ell))g_{\ell}\\
&=&  -\sum_{|\ell| \geq N} \gamma_{X}(\ell)g_{\ell}-\sum_{|\ell| < N}(\tilde{\gamma}_{X}(\ell)-\E\tilde{\gamma}_{X}(\ell))g_{\ell}\\
&=& -T_{1}+ T_{3}
\end{eqnarray*}
\begin{eqnarray*}
T_{1}&=& \sum_{|\ell| \geq N} \gamma_{X}(\ell)g_{\ell}\\
T_{3} &=& \sum_{|\ell| < N}(\tilde{\gamma}_{X}(\ell)-\E\tilde{\gamma}_{X}(\ell))g_{\ell}
\end{eqnarray*}
\begin{eqnarray*}
\|T_{1}\|^{2}_{\mathcal{H}\prime_{s}}&\leq & \sum_{|\ell| \geq N} (1+|\ell|)^{-s}\gamma_{X}(\ell)^{2} \leq \frac{1}{N}\sum_{|\ell| \geq N} \gamma_{X}(\ell)^{2}=\frac{\gamma}{N}<\infty\\
\E\|T_{3}\|^{2}_{\mathcal{H}\prime_{s}}&\leq& \sum_{|\ell| < N} (1+|\ell|)^{-s}\E (\tilde{\gamma}_{X}(\ell)-\E\tilde{\gamma}_{X}(\ell))^{2} \leq \sum_{|\ell| < N} (1+|\ell|)^{-s} \var(\tilde{\gamma}_{X}(\ell))
\end{eqnarray*}
\begin{eqnarray*}
\|J_{N}(g)-J(g)\|^{2}_{\mathcal{H}\prime_{s}}&\leq & 2(\|T_{1}\|^{2}_{\mathcal{H}\prime_{s}}+\E\|T_{3}\|^{2}_{\mathcal{H}\prime_{s}} )\\
&=& \frac{2}{N}(\gamma+ \sum_{|\ell| < N} (1+|\ell|)^{-s} (\kappa_{4}+2\gamma))\\
&=& \frac{2}{N}(\gamma+ c_{s} (1+|\ell|)^{-s} (\kappa_{4}+2\gamma))
\end{eqnarray*}
\item
\begin{eqnarray*}
\sigma_{m}^{2} &=& n\E(\sum_{|\ell| \leq m}(\tilde{\gamma}_{X, N}(\ell)- \gamma_{X}(\ell))\hat{g}(\ell ))^{2}\\
&=& n\var(\sum_{|\ell| \leq m}(\tilde{\gamma}_{X, N}(\ell)- \gamma_{X}(\ell))\hat{g}(\ell ))\\
&=& n \sum_{|\ell| \leq m} \cov\biggl( \big((\tilde{\gamma}_{X, N}(\ell)- \gamma_{X}(\ell))\hat{g}(\ell )\big),\big((\tilde{\gamma}_{X, N}(k)- \gamma_{X}(k))\hat{g}(k)\big)\biggl)
\end{eqnarray*}
Let $A(k)=(\tilde{\gamma}_{X, N}(k)- \gamma_{X}(k)) $ and $A(\ell)=(\tilde{\gamma}_{X, N}(\ell)- \gamma_{X}(\ell))$, then

\begin{equation*}
\sigma_{m}^{2} = n \sum_{|\ell| \leq m} \cov(A(k), A(\ell))\hat{g}(k )\hat{g}(\ell )
\end{equation*}
\item
\begin{eqnarray*}
\sigma_{m}^{2}-\sigma^{2} &\leq & 2|\hat{g}|_{\mathcal{H}_{s}} \quad|\hat{g}-\hat{g}_{m}|_{\mathcal{H}_{s}}\sqrt{B}\\
B &=& \sum_{|k|>m} (1+|k|)^{2s} \var(A(k)) \sum_{\ell} (1+|\ell|)^{2s} \var(A(\ell)) <\infty
\end{eqnarray*}

\end{enumerate}

\noindent Let $g:\mathcal{H}'_{s}\to E_{k}$, where $g$ denotes the orthogonal projection on the closed linear subspace $E_{k} \subset \mathcal{H}'_{s}$, generated by $(e_{\ell})_{|\ell| geq L}$ with $e_{\ell}(\lambda)=e^{i\ell\lambda}$.
\\
Suppose that $g \in E_{k}= \{g_{i}=0, |i| \geq k \}$. Then
\begin{equation*}
\sqrt{N}(J_{X, N}(g)-J_{X}(g) ) \limiteloi N(0, \sigma^{2}(g))
\end{equation*}
because is equal to $\sum_{|i| \leq k} g_{i}Z_{N(i)}$.
\\
Also note that if $g \in \mathcal{H}'_{s}$, and $g^{k}$ is the projection on $E_{k}$, then $\sigma^{2}(g^{(k)})\to \sigma^{2}(g)$ and therefore
\begin{equation*}
\sqrt{N}[\tilde{J}_{X, N}(g)-J_{X}(g)] \limiteloi N(0, \sigma^{2}(g)) \, \, \forall g \in \mathcal{H}'_{s}.\qquad\square
\end{equation*}
\section{Examples}

We present here examples where weak dependence, as defined in Section \ref{sec:dep}, holds. First, we focus on some general
classes of processes before and then we will study specific models.

A generic sample follows

\begin{Def}
Let $(\epsilon_t)_{t \in \Z}$ be a sequence of real-valued random variables and let $F: \R^\Z \to E$ be a measurable function.
The sequence $(X_t)_{t \in \Z}$ defined by
\begin{equation}
    \label{eq:bern-shi}
X_{t} = F(\epsilon_{t-j} ; j \in \Z)
\end{equation}
is called a Bernoulli shift.
\end{Def}

The class of Bernoulli shifts is very general. It provides examples of processes that are weakly dependent but not mixing
(see \cite{Ros85}).

\subsection{Markov processes}

Markov processes can be represented as Bernoulli shifts. Consider an $\R^D$-valued Markov process, driven by the recurrence
equation

\begin{equation}
\label{eq:mark-rec} X_t = f(X_{t-1}, \epsilon_t) \qquad (t\in \Z)
\end{equation}

for some i.i.d. sequence $(\epsilon_t)_{t\in \Z}$ with $\E(\epsilon_0) = 0$, $\epsilon_t$ independent of $\{X_s ; s <t\}$ and $f : \R^D
\times \R \to \R^D$. Then the function $F$ in (\ref{eq:bern-shi}) is defined implicitly (if it exists) by the relation

\begin{equation*}
F(x) = f(F(x'), x_{0})
\end{equation*}

where $x = (x_{0},x_{1}, x_{2} \dots)$ and $x'= (x_{1}, x_{2}, x_{3}, \dots)$.

Assume now in representation (\ref{eq:mark-rec}) that $x_0$ is independent of the sequence $(\epsilon_t)_{t\in \N}$. Suppose that,
for some $0 \leq c_i < 1$,

\begin{equation}
\label{eq:duflo-cond} \E|f (0, \epsilon_1)| < \infty \mbox{ and }  \E|f (u,\epsilon_1) -f (v,\epsilon_1)| \leq \sum ^d_{i=1} c_i |u_i
-v_i|,
\end{equation}
\begin{equation*}
c = \sum^d_{i=1} c_i <1 \mbox{ for all } u, v \in  \R^D.
\end{equation*}

Under condition (\ref{eq:duflo-cond}) the Markov process $(X_t)_{t\in \N}$ has a stationary
distribution $\mu$ with finite first moment. Assume now in addition that $x_{0}$ is distributed with $\mu$, that is, the
Markov chain is stationary. Then, if (\ref{eq:duflo-cond}) holds, such a Markov chain is $\theta$-weak dependent with
$\theta_r = c^r \E|x_0|$.

\subsubsection{Nonparametric AR model.}

Consider the real-valued functional (nonparametric) autoregressive model

\begin{equation}
X_t = r(X_{t-1}) + \epsilon_t ,
\end{equation}

where $r: \R \to \R$ and $(\epsilon_t)_{t \in \Z}$ as in (\ref{eq:mark-rec}). This a special example of a Markov process as in
(\ref{eq:mark-rec}). Assume that $|r(u)-r(u')| \leq c|u- u'|$ for all $u, u' \in \R$ and for some $0 \leq c
<1$, and $E|\epsilon_0| < \infty$. Then (\ref{eq:duflo-cond}) with $D = 1$ holds and implies $\theta$-weak dependence with
$\theta_r = c^r\E|Z_0|$.

Here is important to note that the marginal distribution of the innovations $\epsilon_t$ can be discrete. In such a case, classical
mixing properties can fail to hold. For example, consider the simple linear AR(1) model,

\begin{equation*}
X_t = \phi X_{t-1} + \epsilon_t = \sum _{j\geq 0} \phi^j \epsilon_{t-j}, \quad |\phi| < 1.
\end{equation*}

Let $(\epsilon_t)_{t \in \Z}$ be a sequence of i.i.d. Bernoulli variables with parameter $s = \P[\epsilon_t = 1] = 1-\P[\epsilon_t =
0]$. The AR(1) process $(X_t )_{t\in \Z}$ with innovations $(\epsilon_t )_{t\in \Z}$ and AR parameter $\epsilon \in ]0,
\frac{1}{2}]$, is $\theta$-weak dependent with $\theta_r = \phi^r \E|X_0|$, but it is known to be non-mixing. In this context concentration holds. For example, $X_t$ is uniform if $s = \frac{1}{2}$ and it has a Cantor marginal distribution if $s = \frac{1}{3}$ . Hence, without a regularity condition on the marginal distribution of $\epsilon_0$,
Bernoulli shifts or Markov processes may not be mixing.

\subsubsection{Nonparametric ARCH model.} Consider the real-valued functional (nonparametric) ARCH model

\begin{equation*}
X_t = s(X_{t-1})\epsilon_t ,
\end{equation*}

where $s: \R\to \R^+$ and $(\epsilon_t)_{t\in \Z}$ is defined as in (\ref{eq:mark-rec}) with $E|\epsilon_0|^2 = 1$. This is a special example of
a Markov process as in (\ref{eq:mark-rec}) with $f(u,v) = s(u)v$. Assume that $|s(u)-s(u')| \leq c|u-u'|$ for all $u,
u' \in \R$ and for some $0 \leq c < 1$. Then (\ref{eq:duflo-cond}) with $D = 1$ holds and implies $\theta$-weak
dependence with $\theta_r = cr\E|X_0|$. Again, the innovation distribution is allowed to be discrete.

\subsubsection{Nonparametric AR-ARCH model.}

We interest now in the combination of AR and ARCH models. This new process have nonparametric conditional mean and variance structure,

\begin{equation*}
X_t = r(X_{t-1}) + s(X_{t-1})\epsilon_t ,
\end{equation*}

with $r(\cdot), s(\cdot)$ and $(\epsilon_t)_{t \in \Z}$ as in the examples above. Assume the Lipschitz conditions on
$r(\cdot)$ and $s(\cdot)$ with constants $c_r$ and $c_s$ , respectively. If $c_r+c_s = c < 1$, the process satisfies
$\theta$-weak dependence with $\theta_r = cr\E|X_0|$.

\subsubsection{Bilinear model.}

We consider the simple bilinear process with the following recurrence equation

\begin{equation*}
X_t = aX_{t-1} + bX_{t-1} \epsilon_{t-1} +\epsilon_t,
\end{equation*}

where $(\epsilon_t)_{t\in \Z}$ is as in (\ref{eq:duflo-cond}). Such causal processes are associated with chaotic
representation with stationary
\begin{equation*}
F(u) = \sum^{\infty}_{j=0} u_j  \prod_{s=1}^j (a+b u_s), \qquad u = (u_0, u_1, u_2, \dots).
\end{equation*}

If the process is stationary and $c = \E|a + b \epsilon_0| < 1$, the process satisfies $\theta$-weak dependence, with $\theta_r = \frac{c^r (r +1)}{1-c}$.



\bibliographystyle{apalike}

\end{document}